\documentclass[preprint,twocolumn,superscriptaddress,6pt]{revtex4-1}
\usepackage{amsmath,natbib,graphicx,subfigure,amssymb,graphics,amsmath,mathrsfs,CJK,color}
\usepackage{multirow,fancyhdr,color,bm,tabularx,psfrag,geometry,dcolumn}
\usepackage[dvipdfm,colorlinks=true,linkcolor=blue,urlcolor=blue,citecolor=blue]{hyperref}
\usepackage[left]{lineno}
\def\footnoterule{\kern -1mm \hrule width 6.6cm \kern 2.2mm}%
\usepackage{tikz,xcolor,hyperref}
\definecolor{lime}{HTML}{A6CE39}
\DeclareRobustCommand{\orcidicon}{%
    \begin{tikzpicture}
    \draw[lime, fill=lime] (0,0)
    circle [radius=0.16] node[white]
   {{\fontfamily{qag}\selectfont \tiny ID}};\draw[white, fill=white] (-0.0625,0.095)
    circle [radius=0.007];
    \end{tikzpicture}
    \hspace{-2mm}}
\foreach \x in {A, ..., Z}
{\expandafter\xdef\csname orcid\x\endcsname{\noexpand\href{https://orcid.org/\csname orcidauthor\x\endcsname}{\noexpand\orcidicon}}}

\begin{document}

\title{\LARGE Photoexcited carriers transfer properties in a doped double quantum dots photocell}

\author{Sheng-Nan Zhu}
\affiliation{Department of Physics, Faculty of Science, Kunming University of Science and Technology, Kunming, 650500, PR China}
\affiliation{Center for Quantum Materials and Computational Condensed Matter Physics, Faculty of Science, Kunming University of Science and Technology, Kunming, 650500, PR China}
\author{Shun-Cai Zhao\textsuperscript{\orcidA{}}}
\email[Corresponding author: ]{zsczhao@126.com}
\affiliation{Department of Physics, Faculty of Science, Kunming University of Science and Technology, Kunming, 650500, PR China}
\affiliation{Center for Quantum Materials and Computational Condensed Matter Physics, Faculty of Science, Kunming University of Science and Technology, Kunming, 650500, PR China}
\author{Lin-Jie Chen}
\affiliation{Department of Physics, Faculty of Science, Kunming University of Science and Technology, Kunming, 650500, PR China}
\affiliation{Center for Quantum Materials and Computational Condensed Matter Physics, Faculty of Science, Kunming University of Science and Technology, Kunming, 650500, PR China}
\author{Qing Fang}
\affiliation{Department of Physics, Faculty of Science, Kunming University of Science and Technology, Kunming, 650500, PR China}
\affiliation{Center for Quantum Materials and Computational Condensed Matter Physics, Faculty of Science, Kunming University of Science and Technology, Kunming, 650500, PR China}

\begin{abstract}
Identifying the behavior of photoexcited carriers is one method for empirically boosting their transfer efficiencies in doped double quantum dots (DQDs) photocells. The photoexcited carriers transfer qualities were assessed in this study by the output current, power, and output efficiency in the multi-photon absorption process for a doped DQDs photocell, and an optimization technique is theoretically obtained for this   proposed  photocell model. The results show that some structure parameters caused by doping, such as gaps, incoherent tunneling coupling, and symmetry of structure between two vertically aligned QDs, can remarkably control the photoexcited carriers transfer properties, and that slightly increasing the ambient temperature around room temperature is beneficial to the transfer performance in this doping DQDs photocell model. Thus, our scheme proves a way to optimized strategies for DQDs photocell.
\begin{description}
\item[PACS: ]{78.20.Ci, 42.50.Gy, 81.05.Xj, 78.67.Pt }
\item[Keywords]{ Photoexcited carriers transfer, doped double quantum dots photocell, photovoltaic properties}
\end{description}
\end{abstract}

\maketitle
\section{Introduction}

Because of the quantum confinement effect, it is feasible to readily modify the optical and electrical properties of quantum dots (QDs) by varying their physical parameters (size and shape)\cite{2001Formation,vogel2002quantum,2002On,nozik2009making,Zhao2009,2017General}. As a result, QDs have evolved into adaptive and effective tools for absorbing light and then manipulating, directing, and changing it into various types of energy. QDs are excellent solar absorbers due to their high absorption coefficients and broad, adjustable absorption spectra\cite{2014Noise}. QDs, in particular, have the ability to transform optical light into electrical energy\cite{ZHONG2021104503,PhysRevB.87.035429}. When QDs are combined with a metal contact (which may result in a Schottky junction) or another semiconductor to form a heterostructure, the photoexcited charge in the QD layers can be separated and used to power photovoltaic devices\cite{PhysRevB.87.035429,12-Zhao2019,ZHONG2021104503}.

In the QDs photovoltaic devices\cite{takagahara1992theory,vogel2002quantum,PhysRevB.87.035429,pimachev2016large,rimal2016giant,2017General,12-Zhao2019,2019High}, the flow of energy can easily be tuned by its optical, electrical, physical key parameters, such as its size and shape, internal composition (for example, alloys, core/shell heterostructures, semiconductor/metal interfaces), surface composition (ligand chemistries), and film composition (for example, QD-QD electronic coupling, bulk heterostructure formation, and QD/biological interfaces). In addition, the higher overall conversion efficiency can be achieved via solar cells utilizing the  multiple exciton generation (MEG)\cite{2014Generation,LiMultiple,2022Multiple}, which provides a promising strategy for overcoming the Shockley-Queisser (SQ) limit\cite{1961Detailed} in traditional single-junction solar cells, and MEG process in QDs is proved to be much more efficient than bulk semiconductors by theory and experiments\cite{2014Generation}. Doping technique is confirmed to be a better implementation of MEG and has a stronger photovoltaic response in QD solar cells\cite{pimachev2016large}. Furthermore, the introduction of intermediate energy levels can be achieved by doping techniques and more  photons were absorbed\cite{2019High,Sikder2013Optimization,2020Experimental}. Thus, the energy transfer can be enhanced through impact ionization and the output efficiency was improved\cite{2013Efficiency} compared to the single junction solar cells. In general, doping can be divided into internal and external doping\cite{2018Frontier,2016Photoconductive} in QDs, which usually leads to discrete quantized energy levels in QDs, and in turn absorb photons from different regions\cite{2017Synthesis,2012Theoretical}. The long-lived photogenerated carriers can be generated in doped QDs\cite{hodes2008comparison}, which exhibited a better photovoltaic properties \cite{2000Luminescence,2012Mn,2016Controlled}. In addition, the self-quenching problem can be avoided in the doped QDs\cite{2014Self,Erin2016Self}. However, in the DQDs photocell, there's been little attention to the quantum photoexcited carriers transfer properties dependent some physical parameters caused by the doping techniques so far. Hence, it is significant to investigate the carriers transfer properties dependent some structure parameters caused by doping in the DQDs photocell.

With the quantum photoexcited carriers transfer properties in mind, we propose a doped DQDs photocell model and focus on its carriers transfer properties dependent some physical parameters caused by the doping techniques. And this paper is organized as follows: In Sec. 2, we propose a doped DQDs photovoltaic cell model and its quantum description. In Sec.3, the output photovoltaic characteristics of the DQDs photocell model are discussed in Sec. 3. Finally, the summary of this paper is given in the last section.

\section{THEORETICAL MODEL AND EQUATIONS}
\subsection{Hamiltonian of the DQDs photocell }

\begin{figure}[htp]
\centering
\includegraphics[width=0.95\columnwidth] {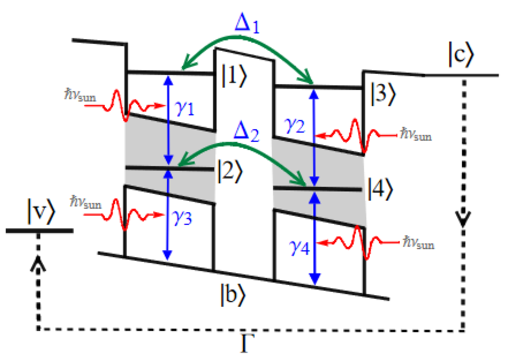}
\caption{(Color online) Schematic diagram of a electronic states of the DQDs photocell. $|c\rangle$ and $|v\rangle$ depict the conduction and valence states of the contacts, modeling a semiconductor where the DQDs are embedded.  Dashed arrow shows the process of relaxation with rate $\Gamma$ from $|c\rangle$ to $|v\rangle$. }
	\label{Fig.1}
\end{figure}

In recent decades, semiconductor QDs have been used as active nanostructured devices for absorbing photons, and the continued progress in nanofabrication technology allowed the fine adjustment of electron and hole band offsets\cite{2006Engineering}. In this paper, we propose two vertically aligned QDs photocell model, which can be implemented in a typical \(InAs/GaAs\) semiconductor QD molecular with different sizes\cite{Zhao2009,2019High}. We focus on the contribution of ion doping technique in the photoexcited carriers transfer properties related to the volt-ampere characteristics, delivering power and conversion efficiency, as indicated in Fig.\ref{Fig.1}.

The doped QDs are depicted by gray areas in the two vertically aligned quantum dots, and their eigenstates ($|2\rangle$ and $|4\rangle$) are represented by the horizontal black solid lines. Their common ground state is denoted by $|b\rangle$, $|1\rangle$ and $|3\rangle$ characterize the excited states (shown in Fig.\ref{Fig.1}). Processes of incoherent tunneling are denoted by the green double arrows between the states $|1\rangle$ and $|3\rangle$, $|2\rangle$ and $|4\rangle$ with $\Delta_{1}$ and $\Delta_{2}$, respectively. Experimentally, the above two parameters depend on the sizes of the barrier between two vertically aligned QDs\cite{2006Engineering}. These nanostructures are embedded in doped semiconductors represented by the conduction $|c\rangle$ and valence $|v\rangle$ reservoir states in the right and left sides of the photocell, which models the action of a p-n junction. The dashed arrow shows the process of relaxation with rate $\Gamma$ from $|c\rangle$ to $|v\rangle$, in which the photocurrent is produced when an electron is released from state $|c\rangle$ resulting in a flow from $|c\rangle$ to $|v\rangle$. The carrier transport process initiated by the absorbed photons with energies $\hbar\nu_{sun}$, as shown by the red wavy line in the Fig.\ref{Fig.1}, drives the carriers at rates $\gamma_{i}(i=1,2,3,4)$ describing the decay rates associated with the electron-hole recombination process linked to the transitions $|1\rangle$ $\leftrightarrow$ $|2\rangle$, $|3\rangle$ $\leftrightarrow$ $|4\rangle$, $|2\rangle$ $\leftrightarrow$ $|b\rangle$, $|4\rangle$ $\leftrightarrow$ $|b\rangle$, respectively. The states of the two vertically aligned QD are coupled to the contacts through phononic coupling at ambient temperature $T_{a}$ with rates $\gamma_{c}$ and $\gamma_{v}$ of the conduction band $|c\rangle$ and valence $|v\rangle$ reservoir, respectively.

According to the above model, the total Hamiltonian of the system is written formally as,

\vskip -0.3cm
\begin{equation}
\hat{H}_{total}=\hat{H}_{s}+\hat{H}_{B}+\hat{H}_{V},
\end{equation}\vskip -0.0cm

\noindent  where $\hat{H}_{s}$=$\hat{H}_{0}+\hat{H}_{i}$, $\hat{H}_{0}$=$\sum_{i}\omega_{i}|i\rangle\langle i|$ (i=b, 1, 2, 3, 4, c, v) is the eigen Hamiltonian of the seven-level system. $\hat{H}_{i}$=$\Delta_{1}|1\rangle\langle 3|+\Delta_{2}|2\rangle\langle 4|+H.c.$ denotes the incoherent tunneling processes between the states $|1\rangle$ and $|3\rangle$, $|2\rangle$ and $|4\rangle$ with their corresponding coupling strengths are $\Delta_{1}$ and $\Delta_{2}$, respectively. The second term, $\hat{H}_{B}$, describes the harmonic oscillator baths represented by,

\vskip -0.3cm\begin{equation}
\hat{H}_{B}= \sum_{i=1}^{4}\sum_{k}\omega_{ik}\hat{a}_{ik}^ {\,\,\dag}\hat{a}_{ik}+\sum_{i=c,v}\sum_{k}\omega_{ik}\hat{b}_{ik}^ {\,\,\dag}\hat{b}_{ik},
\end{equation}\vskip -0.3cm

\noindent  accounts for the ambient environmental reservoirs during photoexcited carriers transport, respectively. Where $\hat{a}_{ik}^ {\dag}(\hat{a}_{ik})$, $\hat{b}_{ik}^ {\dag}(\hat{b}_{ik})$ describes the production (annihilation) operators of the corresponding environmental reservoirs with its k-th noninteracting photon (phonon) mode. Finally, the third term $\hat{H}_{V}=\hat{V}_{1}+\hat{V}_{2}+\hat{V}_{3}+\hat{V}_{4}+\hat{V}_{c}+\hat{V}_{v}$ denotes the Hamiltonian of the interactions between the doped DQDs system and the different ambient reservoirs. Their expressions are respectively written as follows,

\vskip -0.3cm
\begin{eqnarray}
\hat{V}_{1}=\sum_{k}\omega_{1k}(|2\rangle\langle 1|\hat{a}_{1k}^ {\dag}+|1\rangle\langle 2|\hat{a}_{1k}),\nonumber\\
\hat{V}_{2}=\sum_{k}\omega_{2k}(|b\rangle\langle 2|\hat{a}_{2k}^ {\dag}+|2\rangle\langle b|\hat{a}_{2k}),\nonumber\\
\hat{V}_{3}=\sum_{k}\omega_{3k}(|4\rangle\langle 3|\hat{a}_{3k}^ {\dag}+|3\rangle\langle 4|\hat{a}_{3k}),\nonumber\\
\hat{V}_{4}=\sum_{k}\omega_{4k}(|b\rangle\langle 4|\hat{a}_{4k}^ {\dag}+|4\rangle\langle b|\hat{a}_{4k}),\\
\hat{V}_{c}=\sum_{k}\omega_{ck}(|3\rangle\langle c|\hat{b}_{ck}^ {\dag}+|c\rangle\langle 3|\hat{b}_{ck}),\nonumber\\
\hat{V}_{v}=\sum_{k}\omega_{vk}(|b\rangle\langle v|\hat{b}_{vk}^ {\dag}+|v\rangle\langle b|\hat{b}_{vk}).\nonumber
\end{eqnarray}\vskip -0.3cm

\subsection{Master equation}

By tracing out the degrees of freedom of the reservoirs, the dynamic evolution of the doped DQDs photovoltaic cell model in the \(Schr\ddot{o}dinger\) picture can be described by the master equation\cite{1-xu2022differentiation} as follows,

\vskip -0.3cm\begin{equation}
\frac{\partial}{\partial t}\hat{\rho}(t)=-i[\,\hat{H}_{s},\hat{\rho}\,]+\hat{{\mathcal L}}_{A}\hat{\rho}+\hat{{\mathcal{L}}}_{B}\hat{\rho}+\hat{{\mathcal{L}}}_{\Gamma}\hat{\rho},
\end{equation}\vskip -0.3cm

\vskip -0.3cm\begin{eqnarray}
\hat{{\mathcal L}}_{1}\hat{\rho}=\frac{\gamma_{1}}{2}[ (n_{\gamma_{1}}+1)\mathcal{D}[\hat{\sigma}_{21}]\rho+n_{\gamma_{1}}\mathcal{D}[\hat{\sigma}_{21}^{\dag}]\rho],\nonumber\\
\hat{{\mathcal L}}_{2}\hat{\rho}=\frac{\gamma_{2}}{2}[ (n_{\gamma_{2}}+1)\mathcal{D}[\hat{\sigma}_{43}]\rho+n_{\gamma_{2}}\mathcal{D}[\hat{\sigma}_{43}^{\dag}]\rho],\nonumber\\
\hat{{\mathcal L}}_{3}\hat{\rho}=\frac{\gamma_{3}}{2}[ (n_{\gamma_{3}}+1)\mathcal{D}[\hat{\sigma}_{b2}]\rho+n_{\gamma_{3}}\mathcal{D}[\hat{\sigma}_{b2}^{\dag}]\rho],\label{Eq.5}\\
\hat{{\mathcal L}}_{4}\hat{\rho}=\frac{\gamma_{4}}{2}[ (n_{\gamma_{4}}+1)\mathcal{D}[\hat{\sigma}_{b4}]\rho+n_{\gamma_{4}}\mathcal{D}[\hat{\sigma}_{b4}^{\dag}]\rho],\nonumber	
\end{eqnarray}\vskip -0.3cm

\noindent where $\hat{H}_{s}$ is the DQDs Hamiltonian. The superoperator $\hat{\mathcal{L}}$ can be divided into three parts $(\hat{{\mathcal L}}_{A}\hat{\rho},\hat{{\mathcal{L}}}_{B}\hat{\rho},\hat{{\mathcal{L}}}_{R}\hat{\rho})$ to describe the interactions with the photon reservoirs, the thermal phonon reservoir, and the contact point connection part in the doped DQDs photovoltaic cell. The first dissipation term $\hat{{\mathcal L}}_{A}\hat{\rho}$=$\hat{{\mathcal L}}_{1}\hat{\rho}+\hat{{\mathcal L}}_{2}\hat{\rho}+\hat{{\mathcal L}}_{3}\hat{\rho}+\hat{{\mathcal L}}_{4}\hat{\rho}$ describes the exciton transport process in the DQDs interacting with the thermal reservoir in Fig.\ref{Fig.1}, and they are listed in Eg.(\ref{Eq.5}).

In Eg.(\ref{Eq.5}), $\gamma_{i}(i=1,2,3,4)$ is the spontaneous decay rate in the corresponding dissipative process, and the Pauli operators are defined as: $\hat{\sigma}_{ij}$=$|i\rangle\langle j|,\hat{\sigma}_{ij}^{\dag}$=$|j\rangle\langle i|$. Their occupation numbers are $n_{i}$=$\frac{1}{exp[\frac{E_{i}}{K_BT_s}]-1}(i$=$\gamma_{1},\gamma_{2},\gamma_{3},\gamma_{4})$ at the solar temperature, respectively. For convenience, we set the gaps  \(E_{\gamma_{1},\gamma_{3}}\)$=$\(E_{1}\), \(E_{\gamma_{2},\gamma_{4}}\)$=$\(E_{2}\) in the following analysis. Moreover, the specific expression of the sign $\mathcal{D}$ acting on any operator A is $\mathcal{D}[\,\hat A\,]\hat\rho$=$2\hat A\hat\rho\hat A^{\dag}-\hat\rho\hat A^{\dag}\hat A-\hat A^{\dag}\hat A\hat\rho$. Similarly, for the transport process from $|3\rangle$ to $|c\rangle$ and $|b\rangle$ to $|v\rangle$ at rates $\gamma_{c}$, $\gamma_{v}$ are written as follows,

\vskip -0.3cm\begin{widetext}
\begin{eqnarray}
\hat{{\mathcal L}}_{B}\hat{\rho}=\hat{{\mathcal L}}_{c}\hat{\rho}+\hat{{\mathcal L}}_{v}\hat{\rho}=	\frac{\gamma_{c}}{2}[ (n_{\gamma_{c}}+1)\mathcal{D}[\hat{\sigma}_{3c}]\rho+n_{\gamma_{c}}\mathcal{D}[\hat{\sigma}_{3c}^{\dag}]\rho]+\frac{\gamma_{v}}{2}[ (n_{\gamma_{v}}+1)\mathcal{D}[\hat{\sigma}_{bv}]\rho+n_{\gamma_{v}}\mathcal{D}[\hat{\sigma}_{bv}^{\dag}]\rho].
\end{eqnarray}
\end{widetext}\vskip -0.3cm

\noindent where $\hat{\sigma}_{3c}$=$|3\rangle\langle c|$, $\hat{\sigma}_{bv}$=$|b\rangle\langle v|$, and the corresponding average phonon numbers $n_{\gamma_{c}}$=$\frac{1}{exp[\frac{E_{3c}}{K_BT_a}]-1}$, $n_{\gamma_{v}}$=$ \frac{1}{exp[\frac{E_{bv}}{K_B T_a}]-1}$, with the ambient temperature $T_a$. Finally, the last dissipation term in Eq.(1) describes the output electronic current through an external circuit, at a relaxation rate $\Gamma$ from $|c\rangle$ decaying to $|v\rangle$ which is defined as,

\vskip -0.3cm\begin{equation}
\hat{{\mathcal{L}}}_{\Gamma}\hat{\rho} = \frac{\Gamma}{2}(2|v \rangle\langle c|\hat\rho|c \rangle\langle v|-|c \rangle\langle c|\hat\rho-\hat\rho|c \rangle\langle c|).
\end{equation}\vskip -0.3cm

\subsection{Steady-state solution for photovoltaic performance}

Under the Weisskopf-Wigner approximation\cite{1974On}, the elements of the density matrix can be derived and the diagonals are written in the following form,

\vskip -0.3cm\begin{widetext}
\begin{eqnarray}
&&\dot{\rho}_{11}=-i\Delta_{1}(\rho_{31}-\rho_{13})+\gamma_{1}[n_{\gamma_{1}}\rho_{11}-(n_{\gamma_{1}}+1)\rho_{11}],\nonumber\\
&&\dot{\rho}_{22}=-i\Delta_{2}(\rho_{42}-\rho_{24})+\gamma_{1}[(n_{\gamma_{1}}+1)\rho_{11}-n_{\gamma_{1}}\rho_{22}]+\gamma_{3}[n_{\gamma_{3}}\rho_{bb}-(n_{\gamma_{3}}+1)\rho_{22}],\\
&&\dot{\rho}_{33}=i\Delta_{1}(\rho_{31}-\rho_{13})+\gamma_{2}[n_{\gamma_{2}}\rho_{44}-(n_{\gamma_{2}}+1)\rho_{33}]+\gamma_{c}[n_{\gamma_{c}}\rho_{cc}-(n_{\gamma_{c}}+1)\rho_{33}],\nonumber\\
&&\dot{\rho}_{44}=i\Delta_{2}(\rho_{42}-\rho_{24})+\gamma_{2}[(n_{\gamma_{2}}+1)\rho_{33}-n_{\gamma_{2}}\rho_{44}]+\gamma_{4}[n_{\gamma_{4}}\rho_{bb}-(n_{\gamma_{4}}+1)\rho_{44}],\nonumber\\
&&\dot{\rho}_{cc}=\gamma_{c}[(n_{\gamma_{c}}+1)\rho_{33}-n_{\gamma_{c}}\rho_{cc}]-\Gamma\rho_{cc},\nonumber\\
&&\dot{\rho}_{vv}=\Gamma\rho_{cc}+\gamma_{v}[n_{\gamma_{v}}\rho_{bb}-(n_{\gamma_{v}}+1)\rho_{vv}],\nonumber
\end{eqnarray}
\end{widetext}\vskip -0.2cm

\noindent And the off-diagonal elements are as follows,

\vskip -0.3cm\begin{widetext}
\begin{eqnarray}	
&&\dot{\rho}_{13}=-i\rho_{13}(\omega_{1}-\omega_{3})-i\Delta_{1}(\rho_{33}-\rho_{11})-\frac{1}{2}\rho_{13}[\gamma_{1}(n_{\gamma_{1}}+1)+\gamma_{2}(n_{\gamma_{2}}+1)+\gamma_{c}(n_{\gamma_{c}}+1)],\nonumber\\
&&\dot{\rho}_{31}=i\rho_{31}(\omega_{1}-\omega_{3})+i\Delta_{1}(\rho_{33}-\rho_{11})-\frac{1}{2}\rho_{31}[\gamma_{1}(n_{\gamma_{1}}+1)+\gamma_{2}(n_{\gamma_{2}}+1)+\gamma_{c}(n_{\gamma_{c}}+1)],\\	&&\dot{\rho}_{24}=-i\rho_{24}(\omega_{2}-\omega_{4})-i\Delta_{2}(\rho_{44}-\rho_{22})-\frac{1}{2}\rho_{24}[\gamma_{1}n_{\gamma_{1}}+\gamma_{2}n_{\gamma_{2}}+\gamma_{4}(n_{\gamma_{4}}+1)+\gamma_{3}(n_{\gamma_{3}}+1)],\nonumber\\	&&\dot{\rho}_{42}=i\rho_{42}(\omega_{2}-\omega_{4})+i\Delta_{2}(\rho_{44}-\rho_{22})-\frac{1}{2}\rho_{42}[\gamma_{1}n_{\gamma_{1}}+\gamma_{2}n_{\gamma_{2}}+\gamma_{4}(n_{\gamma_{4}}+1)+\gamma_{3}(n_{\gamma_{3}}+1)],\nonumber
\end{eqnarray}
\end{widetext}\vskip -0.3cm

Let's consider the case where $|c\rangle$ and $|v\rangle$ are connected by the load (i.e. resistor), the voltage $V$ across the DQDs photocell model can be expressed in terms of the populations of the levels $|c\rangle$ and $|v\rangle$ (according to the Fermi-Dirac statistic) connected to the load as\cite{12-Zhao2019,1-xu2022Photosynthetic}: $eV $=$E_{c}-E_{v}+K_{B}T_{a}\ln\frac{\rho_{cc}}{\rho_{vv}}$, where $E_{i}$ ($E_{i=c,v}$) is the state energy with an occupation $\rho_{ii}$ calculated in the steady state regime, and the effective current of the external circuit is expressed as $j=e\Gamma{\rho}_{cc}$, with $e$ being the fundamental charge. The power delivered by the DQDs photocell to the load is $P=jV$ owing to the incident solar radiation power $P_{s}$=$j\frac{(E_{1}+E_{2}+2(\omega_1-\omega_3))}{e}$\cite{PhysRevA.84.053818,12-Zhao2019}. Thus, the photovoltaic efficiency can be expressed as,

\vskip -0.3cm\begin{equation}
\eta=\frac{E_{c}-E_{v}+K_{B}T_{a}\ln\frac{\rho_{cc}}{\rho_{vv}}}{(E_{1}+E_{2}+2(\omega_1-\omega_3))}.
\end{equation}\vskip -0.3cm

\section{Results and discussions}

In this work, the significant change is that a variety of different gaps will be introduced to this doped DQDs photocell model, which will cause more absorbed photons with different energies. The relationship between the photovoltaic performance and absorbed photons is a  matter of concern. As a result, the photovoltaic properties of the doped DQDs photocell model will be examined in steady state. In addition, certain standard parameters should be chosen before to the study. Spontaneous decay rates in the photons transition process are $\gamma_{1}$=$\gamma_{2}$=$\gamma$, $\gamma_{3}$=$\gamma_{4}$=$10\gamma$, $\gamma_{5}$=100$\gamma$, $\Gamma$=$0.12\gamma$ with $\gamma$ being the scaling parameter. The ambient temperature is set as $T_a$=0.026$eV$\cite{12-Zhao2019} and the temperature of the sun is $T_S$=$0.5eV$\cite{PhysRevA.84.053818} with $K_B$=$1$.

\begin{figure}[htp]
\centering
\includegraphics[width=0.5\columnwidth] {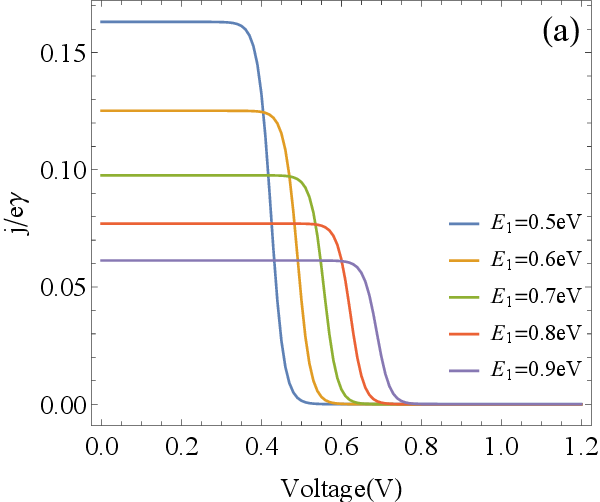}\includegraphics[width=0.5\columnwidth] {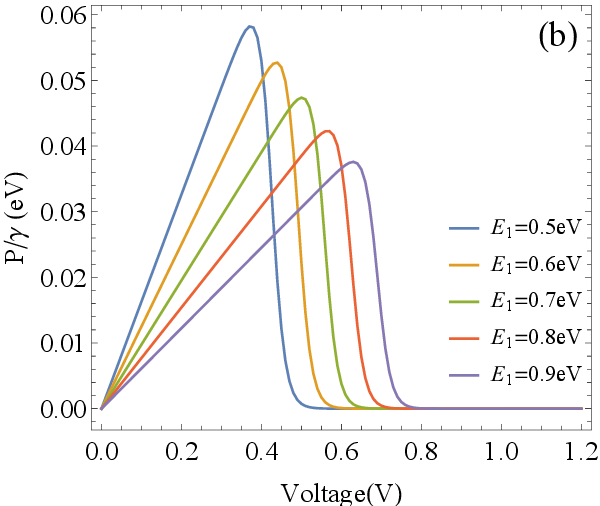}
\hspace{0in}%
\includegraphics[width=0.5\columnwidth] {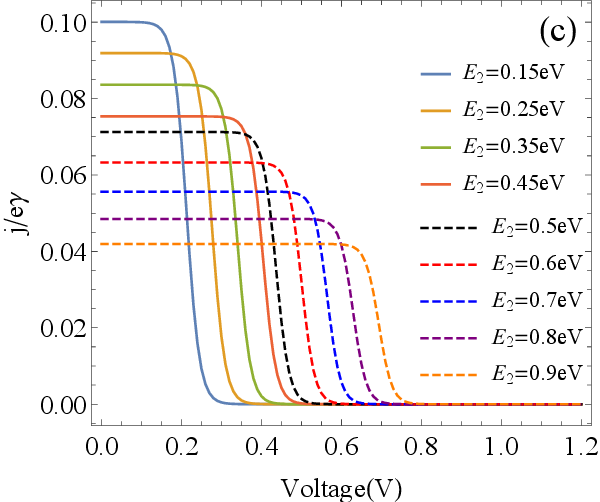}\includegraphics[width=0.5\columnwidth] {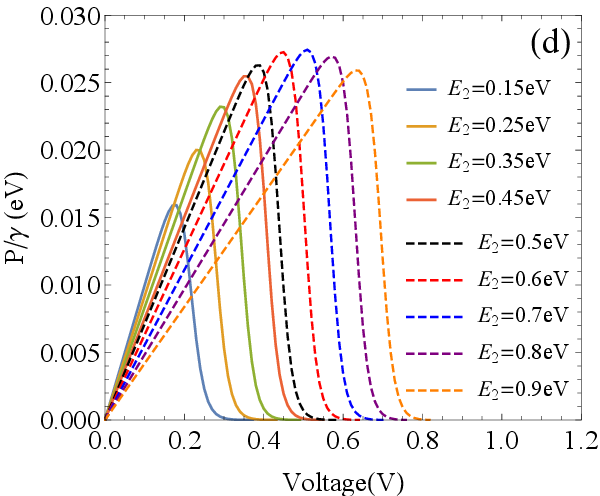}
\caption{(Color online)  Current j (a) and the corresponding output power P (b) as a function of voltage V with different band gaps $E_{1}$.  It takes other parameters are: $\omega_{1}-\omega_{3}$=0.02eV, $\omega_{2}$=$\omega_{4}$, $E_{c}-E_{v}$=$E_{1}$, $E_{2}$=0.65eV, $E_{3c}$=$E_{bv}$=0.25eV, $\Delta_{1}$=3$\Delta_{2}$, $\Delta_{2}$=$4\times 10^{-4}$, $\gamma_{6}$=$100\gamma$.  Current j (c) and the corresponding output power P (d) as a function of voltage V with different band gaps $E_{2}$. Other parameters used are shown below: $\omega_{1}-\omega_{3}$=0.02eV, $\omega_{2}$=$\omega_{4}$, $E_{c}-E_{v}$=$E_{2}$, $\gamma_{6}$=100$\gamma$, $E_{1}$=0.85eV, $E_{3c}$=$E_{bv}$=0.25eV, $\Delta_{1}$=1.5$\Delta_{2}$, $\Delta_{2}$=$4\times 10^{-4}$ with $\gamma$ being the scaling parameter.} \label{Fig.2}
\end{figure}

Ref.\cite{2021The} showed that the optical response could be affected by the impurity locations in QDs owing to the different gaps generated by different locations. The numerous intermediate bands with varying gaps between the conduction and valence bands are similar to the impurity locations in a QD structure, which provides the fact for simulating multi-band gap structures. Therefore, the influence of multi-band gap caused by doping on the photoexcited carriers transfer performance will be discussed in what follows in this paragraph. In this proposed DQDs photocell model, two intermediate bands $|2\rangle$ and $|4\rangle$ will be generated via doping in the two vertically aligned QDs, and its photoexcited carriers transfer properties may be affected by the gaps \(E_{1}\) and \(E_{2}\), respectively. The photovoltaic properties are shown in Fig.\ref{Fig.2} by the current j and the corresponding output power P dependent different gaps \(E_{1}\), \(E_{2}\). And the results indicate different roles in photoexcited carriers transport performance of the two vertically aligned QDs. The decreasing photoexcited carriers transport performance appears when the bandgap energy is greater than 0.5 eV in the left QD, while in the right QD, the narrow gap can improve the output of the system's photoexcited carriers transport performance, and the decreasing carriers transport performance appears when the gap energy is greater than 0.7 eV. From the above, it can be seen that the critical gaps for decreasing carrier transport are not the same in the left and right QDs.

\begin{figure}[htp]
\centering
\includegraphics[width=0.5\columnwidth] {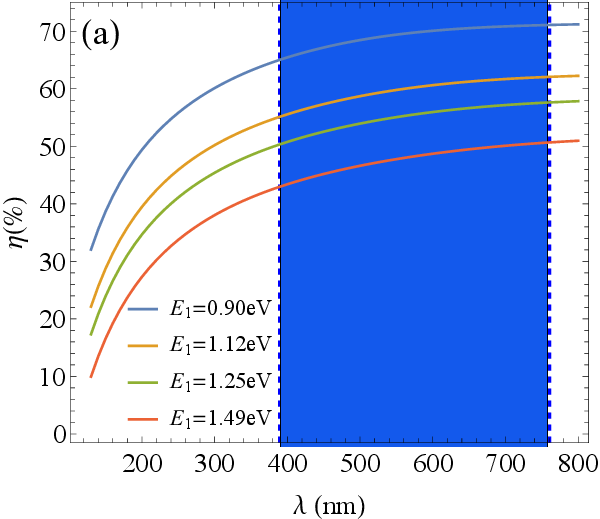}\includegraphics[width=0.5\columnwidth] {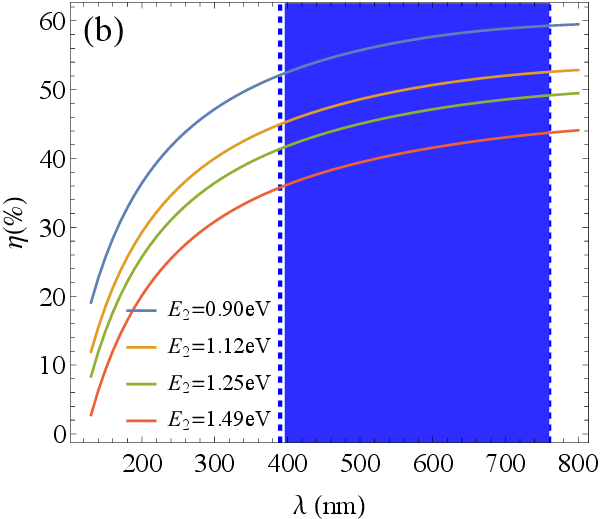}
\hspace{0in}%
\caption{(Color online) Photovoltaic conversion efficiency $\eta$ as a function of the absorbed spectrums $\lambda$ with $E_{2}$=0.65eV in (a), and $E_{1}$=0.95eV in (b). The blue areas corresponds to an absorption spectrum ranging in [390nm, 760nm]. $\omega_{1}-\omega_{3}$=0.02eV, $\omega_{2}$=$\omega_{4}$, $E_{c}-E_{v}$=1.4eV, $E_{3c}$=$E_{bv}$=0.08eV, $\gamma_{6}$=0.05$\gamma$. Other parameters are the same to those in Fig.\ref{Fig.2}(d).}\label{Fig.3}
\end{figure}

Photovoltaic conversion efficiency $\eta$ dependent the absorbed photons spectrum was shown in Fig.\ref{Fig.3} with different gaps $E_{1}$, $E_{2}$, which further proves the conclusion drawn from Fig.\ref{Fig.2}. The blue zones in Fig.\ref{Fig.3} indicate the photovoltaic conversion efficiency $\eta$ in visible region, and they demonstrate the decreasing relation with the increasing gaps when $E_{1}$, $E_{2}$ are greater than 0.9 eV.

\begin{figure}[htp]
\centering
\includegraphics[width=0.6\columnwidth] {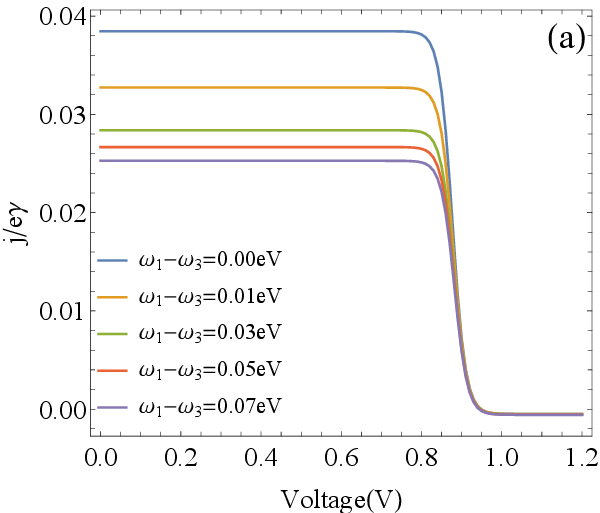}\hspace{0in}\includegraphics[width=0.6\columnwidth] {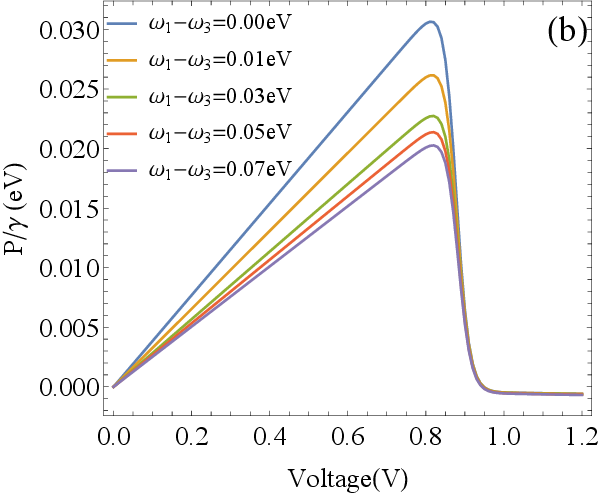}\hspace{0in}\includegraphics[width=0.6\columnwidth] {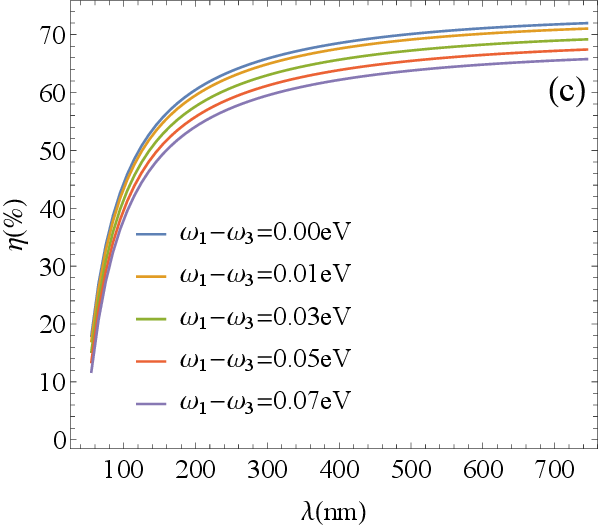}
\hspace{0in}%
\caption{(Color online) Current j (a) and output power P (b) as a function of voltage V with different $\omega_{1}-\omega_{3}$,  and photovoltaic conversion efficiency $\eta$  versus wavelength $\lambda$ with different $\omega_{1}-\omega_{3}$ with $E_{3c}$=$E_{bv}$=0.005eV, $\Delta_{1}$=1.5$\Delta_{2}$, $\Delta_{2}$=$4\times 10^{-4}$. Other parameters are the same to those in Fig.\ref{Fig.3}.}
\label{Fig.4}
\end{figure}

On the other hand, the fabrication process may bring about the problem about the symmetry in two vertically aligned QDs, and its influence on the photoexcited carriers transfer is also an issue worthy of attention. In the following, we will simulate the structural features by some physical parameters for this proposed DQDs photocell. Fig.\ref{Fig.4} describes the symmetry or asymmetry features via the difference in gaps between two QDs.
Because in our proposed DQDs photocell model, two vertical QDs have a common ground state $|b\rangle$, and the energy of the ground state $|b\rangle$ is set as zero, $\hbar$=1, the symmetry of the two QDs can be characterized by the gap difference $\omega_{1}-\omega_{3}$. The results show that the two symmetrically aligned QDs are more favorable for the photoexcited carriers transport. The curves in Fig.\ref{Fig.4} (a) indicate that short-circuit current reaches its maximum value when the DQDs are symmetrically aligned, that is, $\omega_{1}-\omega_{3}$=0, while the short-circuit current decreases gradually with the increments in $\omega_{1}-\omega_{3}$. The curves for the output power indicate the same physical laws  in Fig.\ref{Fig.4} (b), and the maximum photovoltaic conversion efficiency dependent the wavelength of the absorbed photons was proved in Fig.\ref{Fig.4} (c), when $\omega_{1}-\omega_{3}$=0, i.e., the DQDs are symmetrically aligned.

\begin{figure}[htp]
\centering
\includegraphics[width=0.5\columnwidth] {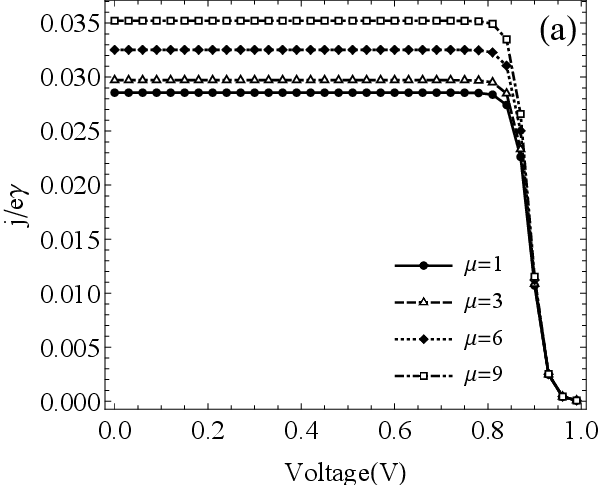}\includegraphics[width=0.5\columnwidth] {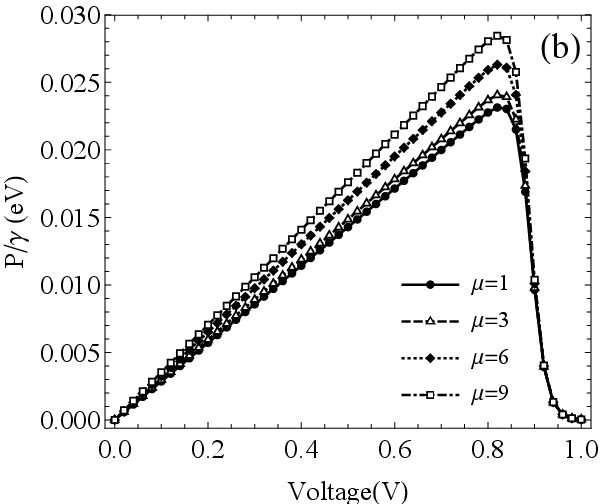}
\hspace{0in}%
\caption{(Color online) Current j and output power P as a function of voltage V with different tunneling coupling ratio $\mu$ in the relation: $\Delta_{1}$=$\mu\Delta_{2}$. The other parameters used are shown below: $E_{3c}$=$E_{bv}$=0.25eV, $E_{1}$=0.95eV, $E_{2}$=0.65eV, $\Delta_{2}$=$2\times 10^{-4}$. Other parameters are the same to those in Fig.\ref{Fig.3}.}
\label{Fig.5}
\end{figure}

In this proposed DQDs photocell model, the doping process may bring out the unequally incoherent coupling coefficients $\Delta_{1}$ (between two excited states $|1\rangle$ and $|3\rangle$) and $\Delta_{2}$ ( between two intermediate bands $|2\rangle$ and $|4\rangle$). Their unequal relationship can be described by the following equation: $\Delta_{1}$=$\mu\Delta_{2}$ with $\mu$ being the tunneling coupling ratio. When the ratio $\mu$ is greater than 1 that tells us that $\Delta_{1}$ is larger than $\Delta_{2}$. The curves in Fig.\ref{Fig.5} clearly demonstrate that a better photoexcited carriers transport is obtained when the incoherent coupling coefficients $\Delta_{1}$ greater than $\Delta_{2}$ in this DQDs photocell system. We argue that the larger incoherent coupling $\Delta_{1}$ between two excited states $|1\rangle$ and $|3\rangle$, can facilitate the carrier transport precess owing to the external output terminal connected the excited state and the ground state. While the incoherent coupling coefficients $\Delta_{2}$ does not leave the carriers directly to the output terminal.

\begin{figure}[htp]
\centering
\includegraphics[width=0.5\columnwidth] {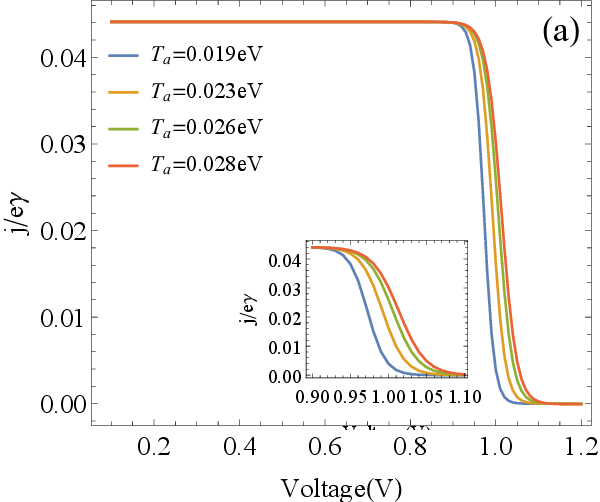}\includegraphics[width=0.5\columnwidth] {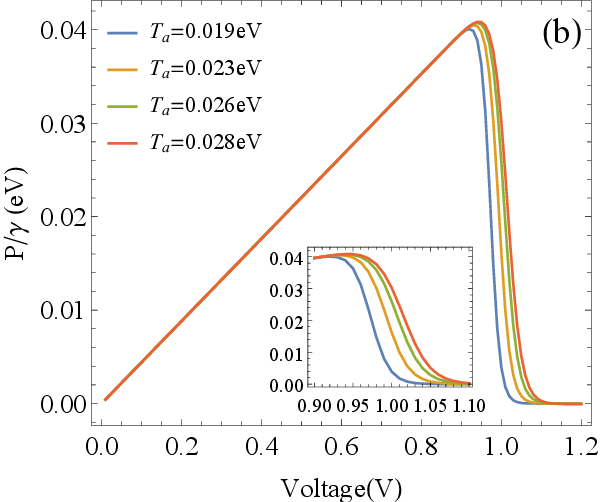}
\hspace{0in}%
\caption{(Color online) Current j and output power P as a function of voltage V with different ambient temperature $Ta$. The other parameters used are shown below: $\omega_{2}-\omega_{4}$=0.02eV, $\gamma_{6}$=100$\gamma$, $\Delta_{1}$=$\Delta_{2}$=$10^{-4}$.  Other parameters are the same to those in Fig.\ref{Fig.5}}\label{Fig.6}
\end{figure}

For any photovoltaic system, the influence of the ambient temperature on its photovoltaic properties is always an unavoidable problem, and an optimization scheme related to the ambient environment temperature is always excited.
The volt-ampere characteristic curves and the output power diagram in Fig.\ref{Fig.6} (a) and (b) display that a slight increment in ambient temperature is beneficial to the photovoltaic performance when this proposed DQDs photocell system is working near the room temperature $T_{a}$=0.0259eV. We speculate that the underlying physics is the rising temperature can accelerate the transport of photogenerated carriers under certain conditions, thus improving the photoexcited carriers transport in this DQDs photocell model.

\section{Conclusion}

In this paper, we used physical parameters to model a doped DQDs photocell system's photoexcited carrier transport mechanism. The effects of bandwidth, incoherent coupling strength between distinct bands in this multiband structure, and ambient temperature on carrier transport behavior were explored, and their related physical regimes were described. The results suggest that the carriers transport performance in this doped DQDs photocell system may be improved. The contributions of the two QDs to the carriers transport process are not same, and the symmetrically developed DQDs system is more advantageous to the carriers transport process. It is recommended to slightly raise the ambient temperature around the room temperature in order to attain better photovoltaic performance. The findings of this research will benefit in the experimental investigation of enhanced performance DQDs photocells.

\section*{Author contributions}

S. C. Zhao conceived the idea. S. N. Zhu performed the numerical computations and wrote the draft, and S. C. Zhao did the analysis and revised the paper. L. J. Chen and Q. Fang gave some discussion in the revised version.

\section{Acknowledgments}

S. C.Zhao is grateful for funding from the National Natural Science Foundation of China (grants 62065009 and 61565008) and Foundation for Personnel training projects of Yunnan Province, China (grant 2016FB009).

\section*{Data Availability Statement}

This manuscript has associated data in a data repository.[Authors' comment: All data included in this manuscript are available upon resonable request by contacting with the corresponding author.]

\bibliography{reference}
\bibliographystyle{unsrt}
\end{document}